# High-throughput Density Functional Perturbation Theory and Machine Learning Predictions of Infrared, Piezoelectric and Dielectric Responses

Kamal Choudhary[1], Kevin F. Garrity[1], Vinit Sharma[2,3], Adam J. Biacchi[4], Angela R. Hight Walker[4], Francesca Tavazza[1]

1 Materials Science and Engineering Division, National Institute of Standards and Technology, Gaithersburg, Maryland 20899, USA.

2 National Institute for Computational Sciences, Oak Ridge National Laboratory, Oak Ridge, TN 37831, USA.

3 Joint Institute for Computational Sciences, University of Tennessee, Knoxville, TN, 37996, USA.

4 Engineering Physics Division, National Institute of Standards and Technology, Gaithersburg, Maryland 20899, USA.

**Abstract**

Many technological applications depend on the response of materials to electric fields, but available databases of such responses are limited. Here, we explore the infrared, piezoelectric and dielectric properties of inorganic materials by combining high-throughput density functional perturbation theory and machine learning approaches. We compute Γ-point phonons, infrared intensities, Born-effective charges, piezoelectric, and dielectric tensors for 5015 non-metallic materials in the JARVIS-DFT database. We find 3230 and 1943 materials with at least one far and mid-infrared mode, respectively. We identify 577 high-piezoelectric materials, using a threshold of 0.5 C/m$^2$. Using a threshold of 20, we find 593 potential high-dielectric materials. Importantly, we analyze the chemistry, symmetry, dimensionality, and geometry of the materials to find features that help explain variations in our datasets. Finally, we develop high-accuracy regression models for the highest infrared frequency and maximum Born-effective charges, and classification models for maximum piezoelectric and average dielectric tensors to accelerate discovery.

# 1 Introduction

The Materials Genome Initiative (MGI)[1] has revolutionized the development of new technologically important materials, which historically has been a time-consuming task that was mainly dominated by trial and error strategies. Since MGI launched in 2011[1,2], high-throughput computational[3-23] and experimental[12,24-27] techniques have become a foundation for the 'materials by design' paradigm. Moreover, the synergy among multidisciplinary sciences, along with rapid advancements in the electronic-structure methods, computational resources, and experimental techniques has made the process of materials design faster and far more efficient in some cases. As a result, the community is gradually migrating towards systematic computation-driven materials selection paradigms,[9,10,13-19,28-35] where functional materials are screened by establishing a direct link between the macroscopic functionality and the atomic-scale nature of the material. We are in a data-rich, modeling-driven era where trial and error approaches are gradually being replaced by rational strategies,[9,36-38] which couple predictions not only from specific electronic-structure calculations of a given property but also by learning from the existing data using machine learning. Subsequent targeted experimental synthesis and validation provide a means of rapid iteration to verify and improve computational models.

Applications of these computational techniques in a high-throughput manner have led to several databases of computed geometries and many physicochemical properties, AFLOW[4], Materials-project[3], Khazana[17], Open Quantum Materials Database (OQMD)[5], NOMAD[7], Computational Materials Repository (CMR)[39], NIMS databases[40] and our NIST-JARVIS databases[6,8,21-23,41-47]. Despite a few systematic experimental databases of IR data (such as https://webbook.nist.gov/chemistry/vib-ser/), a systematic investigation of IR for inorganic materials is still lacking. Similarly, there have been only a few systematic databases developed for Piezoelectric (PZ) and dielectric (DL) materials such as by De Jong et al[48], Petousis et al[49,50], Roy et al.[51], and Choudhary et al[41]. In this work, we significantly expand the scope of these previous efforts by developing systematic databases for infra-red (IR) absorption spectra, piezoelectric tensors, and dielectric tensors.

Vibrational spectroscopy based on infrared (IR), Raman and neutron scattering are ubiquitous methods to probe the chemical bonding, and thus the electronic structure of a material. Infrared frequencies are classified in three categories: far (30-400 cm$^{-1}$), mid (400-4000 cm$^{-1}$) and near (4000-14000 cm$^{-1}$) IR frequencies. Traditionally, IR spectroscopy was only used to probe organic materials, but with the availability of instruments capable to detect frequencies 600 cm$^{-1}$, IR spectroscopy has also been proven successful in distinguishing phases of inorganic compounds[52-54], thermal imaging[55], infrared astronomy[56] and food quality control[57].

The piezoelectric tensor (PZ)[58] describes the change in the polarization of a material due to mechanical stress or strain, or conversely, the change in stress or strain due to an external electric field. Similarly, the dielectric tensor[59,60] describes the change in polarization due to an electric field. A related quantity, the Born-effective charge (BEC) tensor, describes first-order response of the atomic positions to an electric field. All of the PZ tensor-components are zero for materials that have inversion symmetry, leaving 138 space groups with non-zero PZ response while DL tensors and BEC tensors can have non-zero components regardless of symmetry. However, even when non-zero values are allowed, symmetry still strongly restricts the structure of these tensors.

While IR data can be used in infrared-detector design, the associated PZ[61], DL[62] and Born-effective charge data can be utilized for designing sensors, actuators and capacitors[62-64]. In addition, there is significant interest in finding lead-free piezoelectric materials to replace lead zirconate titanate (PZT) in various applications. All the above quantities can be obtained from density functional perturbation theory calculations (DFPT)[65-69]. The DFPT is a well-known technique to effectively calculate the second derivative of the total energy with respect to atomic displacement. The computational phonon-spectroscopy method has been recently shown to be an effective means for characterizing materials, as discussed by Skelton *et. al.*[70] and Kendrik *et. al*[71]. The information obtained using the DFPT method can also be utilized for predicting piezoelectric coefficients, static dielectric matrix and Born effective charges, as described by Gonze *et. al.*[68] and Wu *et. al.*[66]

To validate our calculations, we compare them to the handful of available experimental data and show the uncertainty in predictions. We also identify materials that we predict have exceptional IR, PZ or DL properties, which may be good candidates for experimental synthesis and characterization. Finally, we develop high accuracy supervised machine learning models based on the classification and regression methods to pre-screen high-performance materials without performing additional first-principles calculations. This work is a continuation of our previous datasets for exfoliability[6], elastic[23], optoelectronic[41], topological[21,22], solar-cell efficiency[21], and thermoelectric[43] property, scanning tunneling microscopy image[8] datasets. The complete datasets for IR, PZ, BEC and DL properties are provided on the JARVIS-DFT website (https://www.ctcms.nist.gov/~knc6/JVASP.html ). The JARVIS-DFT is a part of the Materials Genome Initiative (MGI) at NIST.

## 2 Results and discussion

We use density functional perturbation theory (DFPT) to predict the infrared, dielectric and piezoelectric response of insulating materials. Out of 38000 3D materials in the JARVIS-DFT database, we select materials with OptB88vdW bandgaps > 0.1 eV[41] and energy above convex hull < 0.5 eV/atom[72] and in their corresponding conventional cell representation. This leads to 10305 materials. Note that the OptB88vdW gaps are generally underestimated compared to experiments but are useful in pre-screening. The convex hull helps enumerate the thermodynamic stability at 0 K. We further narrow down the list of candidates with the number of atoms in the simulation cell < 20 atoms leading to 7230 systems. We have completed calculations for 5015 materials so far and other calculations are still ongoing. After the DFPT calculations, we obtain the phonon-frequencies at $\Gamma$-point as well. We predict infrared, piezoelectric and dielectric properties of systems only with all positive phonon-frequencies because the materials with negative frequencies are dynamically unstable. Dynamically unstable materials are usually the high symmetry structure of a material that undergoes a symmetry-lowering structural phase transition at low temperatures, such as a ferroelectric transition, but further calculations are necessary to understand these behaviors. For the dynamically stable materials, we also train machine learning models to quickly pre-screen materials for the next set of DFPT calculations. Such machine-learning models have not been reported before to the best of our knowledge. A brief- flow-chart of the whole process is given in Fig. 1. Now we discuss in detail individual components of this work below.

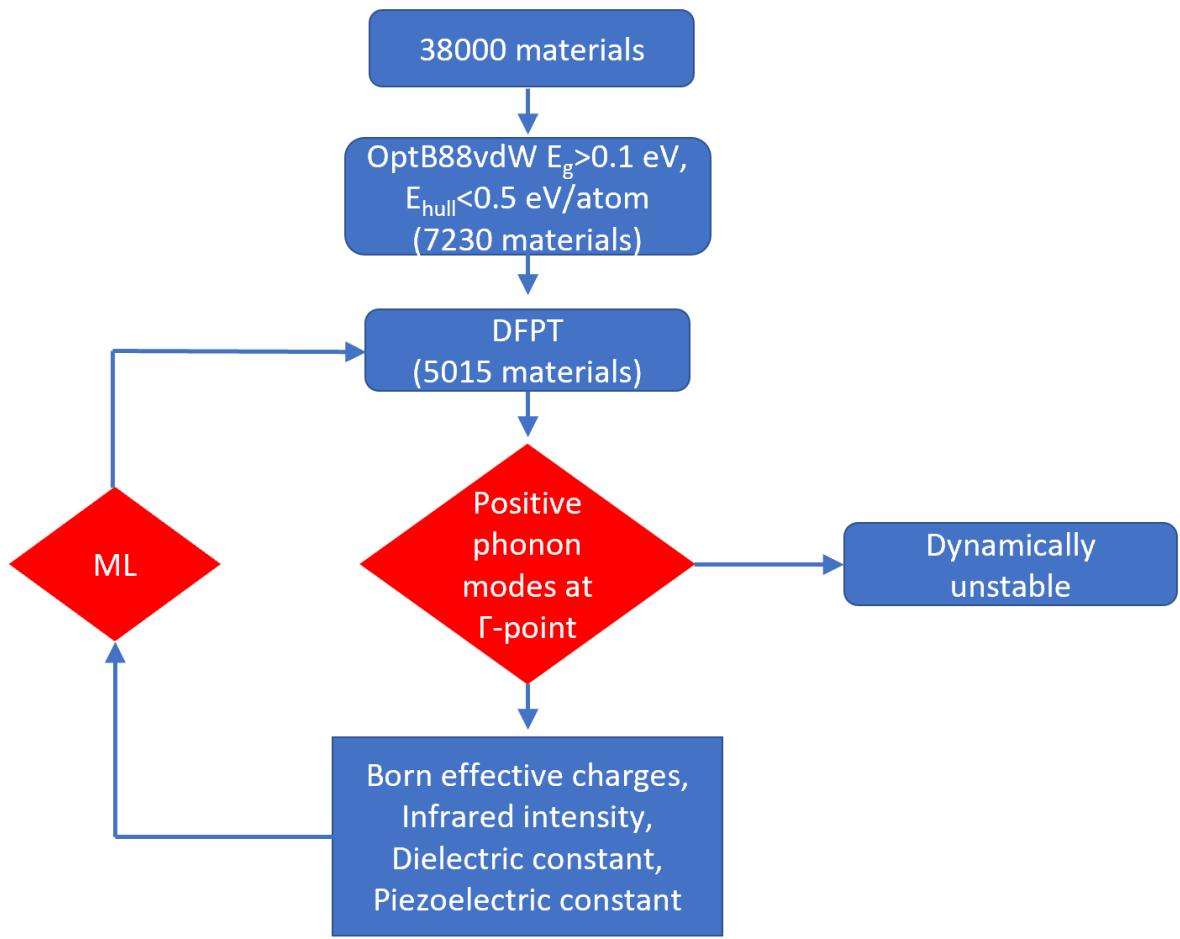

*Fig. 1 Flow-chart portraying different steps for the DFT and ML methods.*

## 2.1 Infrared intensity

As mentioned above, the infrared intensity is important for thermal-imaging, infrared-astronomy, food-quality control. Using Eq. 1-7 (see Methods section) we calculate the IR intensities of 3411 materials with bandgap > 0.1 eV and positive phonon frequencies at Γ-point. We compare nine experimental (Exp.) IR modes with the DFPT results to estimate the data-uncertainty. The comparison is listed in supporting information (see Table S1). Based on the data in table S1, the difference in IR mode location between our calculations and experiments has a mean absolute deviation of 8.36 cm$^{-1}$. This difference is small compared to the total range of IR frequencies, from 0-1000 cm$^{-1}$. Several previous studies also report the difference between DFPT computed frequencies and measured peaks in this range[73-75]. In addition to intrinsic limitations in the DFT-exchange-correlation function, this small discrepancy may be caused by the fact that our calculations are carried out at 0 K while experiments are typically done at 300 K, or that experimental systems can contain defects and impurities, which are absent in our calculations. Given the strong agreement between theory and experiment, we expect that our carefully curated database will be useful for materials characterization.

To analyze the overall trend in the IR data, we plot the results for all materials together in Fig. 2. We observe that most of the modes are less than 1500 cm$^{-1}$ which is consistent with the fact that inorganic materials generally have much lower IR frequencies than organic and soft materials, which typically contain covalently bonded light elements, leading to higher vibrational frequencies. Moreover, a close look at the dataset suggests that 3230 materials have at least one far-IR mode, and 1943 materials have at least one mid-IR mode. As expected, we couldn't find any material with near-IR modes, as the largest frequency we find is 3764 cm$^{-1}$ for $Mg_5(HO_3)_2$ (JVASP-13093). We find 41 elemental systems such as crystalline nitrogen (JVASP-25051), krypton (JVASP-907), xeon (JVASP-25276) etc. have no IR peaks. This can be attributed to the acoustic sum rule (ASR) for Born effective charges, which forces the total Born effective charge of a system to add up to zero. From the dataset, we also observe that 1426 materials have only far-IR, 1804 both far and mid-IR and 139 mid-IR peaks. Some of the materials with the lowest IR frequencies are from the halide family, such as CuBr (JVASP-5176), AgI (JVASP-12023), $TlTe_3Pt_2$ (JVASP-4627), while some of the highest frequencies are obtained for O- or OH-containing compounds, such as $Mg_5(HO_3)_2$ (JVASP-13093). All the materials above 2078 cm$^{-1}$ contain hydrogen and the next highest set of frequencies are obtained for compounds with C-N bonds, which is reasonable, as $\omega \sim m^{-1/2}$ (where $\omega$ is frequency and *m* is the atomic mass). These trends may be useful as a starting point in identifying new materials for infrared-related devices, like detectors, sensors, and lenses.

Next, we compare the DFPT and finite-displacement method (FDM) phonon frequencies (obtained from our elastic-constant database[23]) for 2926 materials and 72624 phonon modes (as shown in Fig. 2b). We find that the DFPT and FDM compare very well (Pearson correlation coefficient of 0.99), and significant differences are only found for a few molecular systems such as crystalline $H_2$, $N_2$ and a few lanthanide oxides (such as ErBiO, JVASP-49979). This consistency suggests an overall high quality of our computational dataset.

In Fig. 2c, we analyze the space-group variation of materials with at least one far-IR (blue) and mid-IR (green) peaks. Some of the high-symmetry space-groups with a high number of materials with far-IR peaks are 36, 80,129, 166 and 216. As shown in Venn diagram in Fig. 2d, we find that a large number of materials with far-IR peaks are chalcogenides (O, S, Se, Te-based compounds); however, our database of inorganic materials is generally biased towards oxides (39% are chalcogenides, 14% are halogens,19% pnictides and 28% others), and we find many examples of halides and pnictides with far-IR peaks as well. Halides such as $MgF_2$, LiF etc. have been used in several astronomical telescopes such as Hubble telescope[76] for infrared astronomy, and non-oxide chalcogenides such as CdTe have been used for infrared night vision cameras. We find that denser materials generally have lower phonon frequencies, which can be observed in the Fig. 2e. This can be explained by the fact that the denser materials have heavier atoms leading to lower phonon frequencies for fixed spring constants. Conversely, high IR modes seem to require low-density materials. Finally, we analyzed how the dimensionality[23] of materials affect the far-IR peaks (Fig.1f). While we find that most of the far IR-peak materials have 3D bonding, we also determine that a significant fraction (20.9%) are low-dimensional, which is slightly higher than their overall representation in the database of 17.2%. The low-dimensional/vdW-bonding is determined based on the bond-topology or lattice constant criteria.[23] Several examples of materials with different dimensionality with low and high-IR modes are shown in Table. 1 and 2. The complete list is provided in the supplementary information (see data-availability section). Identifying low dimensional dielectric materials is important for designing ultrathin flexible electronic devices.

*Table. 1 Some examples of materials with high IR-modes. Complete data is available in the supplementary information.*

| Formula | Dimensionality | JID | Spg. | Max. mode (cm$^{-1}$) |
|---|---|---|---|---|
| **YHO$_2$** | 3D | JVASP-54770 | 11 | 3673.6 |
| **Ca(HO)$_2$** | 3D | JVASP-3714 | 164 | 3669.40 |
| **SrH$_4$O$_3$** | 3D | JVASP-51024 | 6 | 3656.2 |
| **MoH$_4$O$_5$** | 3D | JVASP-54730 | 1 | 3653.8 |
| **SrHClO** | 3D | JVASP-3723 | 186 | 3632.3 |
| **Mg$_5$(HO$_3$)$_2$** | 2D | JVASP-13093 | 164 | 3765.0 |
| **Al$_2$Si$_2$H$_4$O$_9$** | 2D | JVASP-29330 | 1 | 3684.6 |
| **Zn(HO)$_2$** | 2D | JVASP-29800 | 164 | 3645.2 |
| **Al(HO)$_3$** | 2D | JVASP-29385 | 1 | 3576.7 |
| **FePH$_5$CO$_4$** | 2D | JVASP-29727 | 7 | 3410.8 |
| **MoH$_2$Cl$_2$O$_3$** | 1D | JVASP-33602 | 31 | 3582.2 |
| **MnH$_4$(ClO)$_2$** | 1D | JVASP-27170 | 12 | 3446.1 |
| **SnH$_4$(NF)$_2$** | 1D | JVASP-33072 | 12 | 3356.1 |
| **ZnH$_8$(N2Cl)$_2$** | 1D | JVASP-33028 | 12 | 3312.2 |
| **H$_2$C** | 1D | JVASP-33878 | 62 | 2950.5 |
| **BH$_6$CN$_3$F$_4$** | 0D | JVASP-33800 | 160 | 3491.7 |
| **MgH$_4$(ClO)$_2$** | 0D | JVASP-24028 | 12 | 3454.9 |
| **H$_5$NO** | 0D | JVASP-33004 | 19 | 3408.9 |
| **BH$_5$CN$_2$** | 0D | JVASP-33308 | 33 | 3333.1 |
| **MgTe(H$_4$O$_3$)$_3$** | 0D | JVASP-32180 | 146 | 3276.0 |

*Table. 2 Some examples of materials with low IR-modes. Complete data is available in the supplementary information.*

| Formula | Dimensionality | JID | Spg. | Min. mode (cm$^{-1}$) |
| --- | --- | --- | --- | --- |
| BaSr$_2$I$_6$ | 3D | JVASP-50964 | 162 | 3.4 |
| TlSbSe$_2$ | 3D | JVASP-56763 | 4 | 9.4 |
| Ca$_2$SnS$_4$ | 3D | JVASP-40157 | 189 | 9.9 |
| RbCaI$_3$ | 3D | JVASP-36914 | 6 | 5.1 |
| K$_3$ClO | 3D | JVASP-53363 | 38 | 18.8 |
| HgI$_2$ | 2D | JVASP-5224 | 137 | 8.4 |
| Nb$_2$Te$_6$I | 2D | JVASP-5575 | 14 | 14.3 |
| Bi$_2$Se$_3$ | 2D | JVASP-5215 | 62 | 22.5 |
| Ag$_3$SI | 2D | JVASP-29974 | 4 | 28.3 |
| Sb$_2$S$_2$O | 2D | JVASP-30438 | 12 | 29.0 |
| Mn(SbS$_2$)$_2$ | 1D | JVASP-32156 | 12 | 13.9 |
| AlTlSe$_2$ | 1D | JVASP-8210 | 140 | 22.5 |
| NbI$_5$ | 1D | JVASP-5845 | 14 | 24.0 |
| BiSeCl | 1D | JVASP-32942 | 62 | 28.1 |
| SnICl | 1D | JVASP-32820 | 62 | 29.6 |
| P$_2$Se$_5$ | 0D | JVASP-5590 | 14 | 16.3 |
| Ta(TeBr$_3$)$_2$ | 0D | JVASP-5662 | 2 | 25.3 |
| HgIBr | 0D | JVASP-22656 | 36 | 28.4 |
| Ga$_2$PdI$_8$ | 0D | JVASP-5815 | 12 | 30.7 |
| AlTeI$_7$ | 0D | JVASP-5653 | 7 | 32.9 |

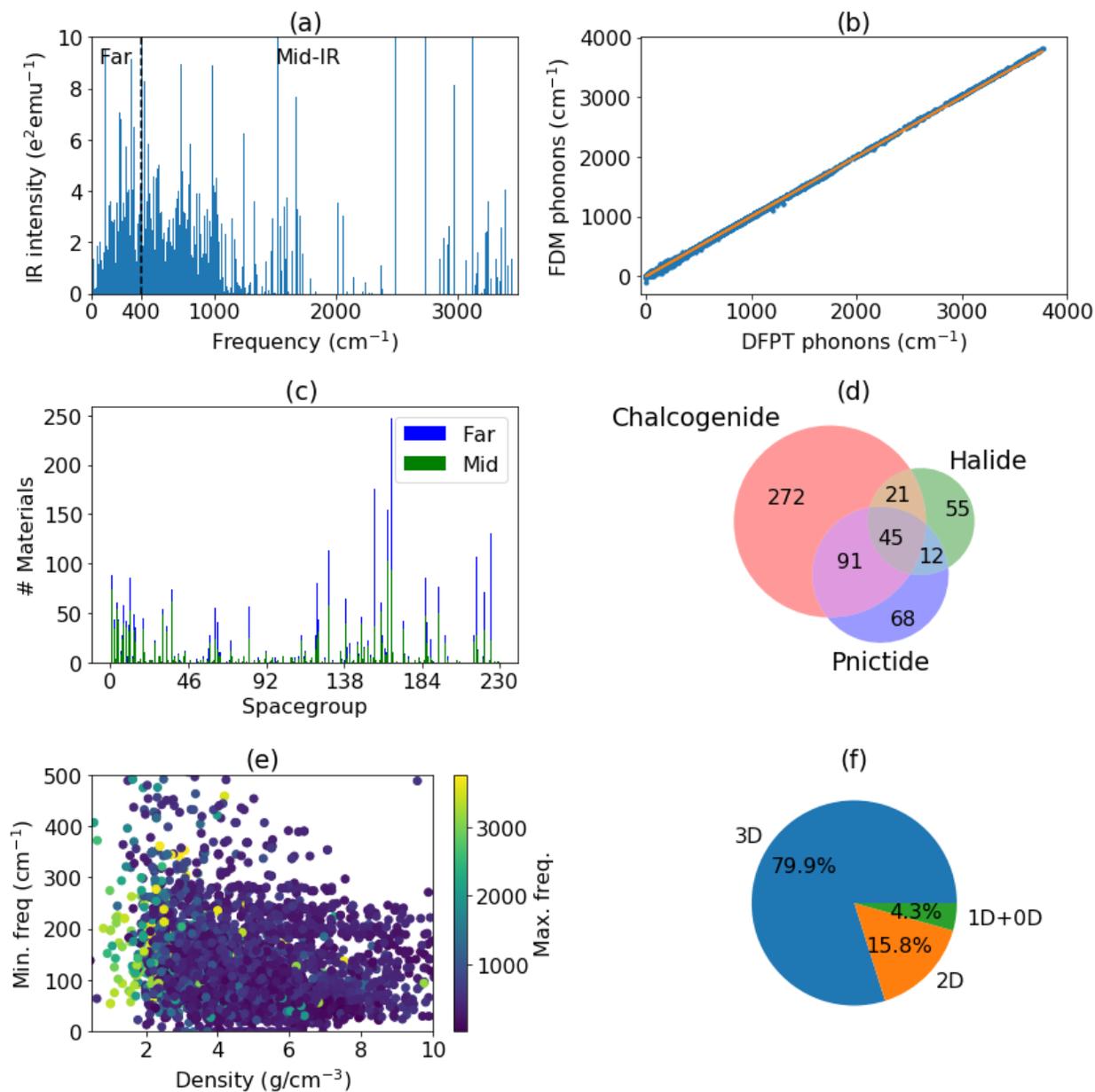

*Fig. 2 Analysis of the IR-data. a) IR peaks for all the materials in the database, b) Comparison of finite-difference (FDM) and DFPT phonon frequencies for conventional cells, c) space-group distributions of materials with at least one far (blue) and mid (green) IR peaks, d) Venn-diagram of the chemistry of materials containing chalcogenides, halides. or pnictides. materials, e) Minimum frequency vs density of the system, f) dimensionality analysis of the far-IR materials.*

## 2.2 Piezoelectric properties

The piezoelectric effect is a reversible process where materials exhibit electrical polarization resulting from an applied mechanical stress, or conversely, a strain due to an applied electric field. Common applications for piezoelectricity include medical applications, energy harvesting devices, actuators, sensors, motors, atomic force microscopes, and high voltage power generation[58]. As discussed in the methods section, PZ responses can be measured under constant strain, giving the piezoelectric stress tensor $e_{ij}$ or constant stress, giving the piezoelectric strain tensor $d_{ij}$. In table S2 we compare DFT computed PZ stress tensors (maximum $e_{ij}$) to experimental ones for several classes of materials, such as oxides, nitrides, and sulfides, and in several crystal structures. We find that the mean absolute deviation in max ($e_{ij}$) is 0.79 C/m$^2$, which is reasonable at least for initial screening purposes. The relaxed-ion contribution to the piezoelectric tensor is proportional to the inverse dynamical matrix (see Eq. 5 in the methods section), which makes this contribution very sensitive to low-frequency IR modes. These modes are often very sensitive to temperature changes and defects, especially in ferroelectric materials, which makes a direct comparison between theory and experiment challenging[77]. In Fig. 3a, we show a histogram of the distribution of the maximum absolute value of the $e_{ij}$ PZ tensor component across the dataset, which shows a peak at zero and a long-tailed distribution. Most of the materials have max $e_{ij}$ less than 1.0, but some of the high PZ materials have max. $e_{ij}$ above 4 C/m$^2$. We choose a 0.5 C/m$^2$ threshold to select whether a material is a good PZ material, giving rise to 577 screened materials. The 0.5 C/m$^2$ threshold value is used in Fig. 3d,3e and 3f. In Fig. 3b we plot a similar histogram for the maximum component of the $d_{ij}$ tensor, the more commonly measured piezoelectric coefficient. The value of $d_{33}$ is generally below 50x10$^{-12}$ C/N as shown in Fig, 3b. Examples of the predicted lead-free high PZ materials, which we do not find reported in literature are: MoO$_3$ (JVASP-30103),

YWN$_3$(JVASP-38813), W(BrO)$_2$ (JVASP-30364), InBiO$_3$(JVASP-34653), NbNO (JVASP-52492), MgTiSiO$_5$ (JVASP-9481), MgCN$_2$ (JVASP-7814), Ca$_2$SnS$_4$ (JVASP-40157), ZnTa$_2$O$_6$ (JVASP-9231), Bi$_3$TaO$_7$ (JVASP-13350). The highest PZ values naturally require a combination of high-Born effective charges and low-frequency IR modes, in combination with a crystal symmetry that allows for the non-zero piezoelectric response. This confluence of factors makes strong piezoelectricity difficult to predict and optimize.

*Table. 3 Some examples of high-piezoelectric coefficient (max-$e_{ij}$) materials with corresponding strain coefficients ($d_{ij}$), if available. Complete data is available in the supplementary information.*

| Material | Dimensionality | JID | Spg | Max($e_{ij}$) C/m$^2$ | Max($d_{ij}$) (x10$^{-12}$ C/N) |
|---|---|---|---|---|---|
| **YWN$_3$** | 3D | JVASP-38813 | 161 | 7.73 | - |
| **RuC** | 3D | JVASP-36402 | 216 | 5.05 | 83.1 |
| **InBiO$_3$** | 3D | JVASP-34653 | 33 | 4.75 | - |
| **NbNO** | 3D | JVASP-52492 | 109 | 4.26 | 132.5 |
| **Ba$_2$Ti$_3$O$_8$** | 3D | JVASP-11504 | 25 | 6.35 | 66.8 |
| **Ta$_2$ZnO$_6$** | 3D | JVASP-9231 | 44 | 3.96 | 49.6 |
| **MgTa$_2$O$_6$** | 3D | JVASP-9226 | 31 | 3.62 | 38.50 |
| **VFeSb** | 3D | JVASP-56856 | 216 | 3.69 | 92.8 |
| **YBiO$_3$** | 3D | JVASP-45986 | 161 | 3.1 | 133.8 |
| **MoO$_3$** | 2D | JVASP-30103 | 7 | 9.44 | 2623.5 |
| **CrHO$_2$** | 2D | JVASP-8621 | 160 | 3.46 | 16.6 |
| **GeTe** | 2D | JVASP-1157 | 160 | 2.76 | 154.6 |
| **SnPSe$_3$** | 2D | JVASP-29622 | 7 | 2.19 | - |
| Mg(SbO$_2$)$_4$ | 2D | JVASP-10736 | 1 | 2.1 | 40.0 |
| V$_2$Pb$_3$O$_8$ | 2D | JVASP-12668 | 5 | 1.75 | - |
| WBr$_4$O | 1D | JVASP-5863 | 79 | 4.04 | - |
| WCl$_4$O | 1D | JVASP-13822 | 79 | 2.02 | 24.7 |
| InGeCl$_3$ | 1D | JVASP-33896 | 160 | 1.45 | 139.0 |
| Br$_2$O | 1D | JVASP-12916 | 33 | 0.94 | 35.7 |
| InSnCl$_3$ | 1D | JVASP-33897 | 160 | 0.68 | - |
| MgTe(H$_4$O$_3$)$_3$ | 0D | JVASP-32180 | 146 | 0.74 | - |
| AsF$_3$ | 0D | JVASP-24657 | 33 | 0.68 | - |
| SeBr | 0D | JVASP-22664 | 41 | 0.66 | 112.43 |
| ClO$_3$ | 0D | JVASP-12460 | 9 | 0.53 | - |
| BrO$_2$F | 0D | JVASP-31165 | 9 | 0.45 | |

As depicted in Fig. 3c we compare the stress and strain-based PZ. We find that there is no obvious relation between the stress and strain PZ tensors, implying that accurate compliance tensor is essential to predict the $d_{ij}$ values. It is important to note that PZ strain tensors are the components which are generally measured during the experiments. Next, in Fig. 3d we analyze the chemistry of high PZ materials (using 0.5 C/m$^2$ as a threshold). Like the IR data, the high PZ materials are dominated by chalcogenides, with very few halides. We find most of the materials with high PZ are 3D, with low-dimensional materials under-represented. Some examples of 3D and low-dimensional materials are given in Table. 3. Analyzing the crystal systems of the high-PZ materials, we observe that orthorhombic systems are the most represented crystal system (Fig. 3e), while space group 216 is the most common space group. As the PZ is a tensor quantity, we analyze the tensor component distribution for the whole dataset in Fig. 4a. All the piezoelectric tensor components have similar distributions with outlier materials with high coefficients. The $e_{33}$ component has the largest number of high-value piezoelectric response materials. This stands in contrast to other tensors, like the elastic tensor, where diagonal components tend to dominate, and many components are almost always near zero. Our database can also be used to identify unusual piezoelectric mechanisms, such as negative piezoelectricity[78], but such detailed analysis is beyond the scope of current work.

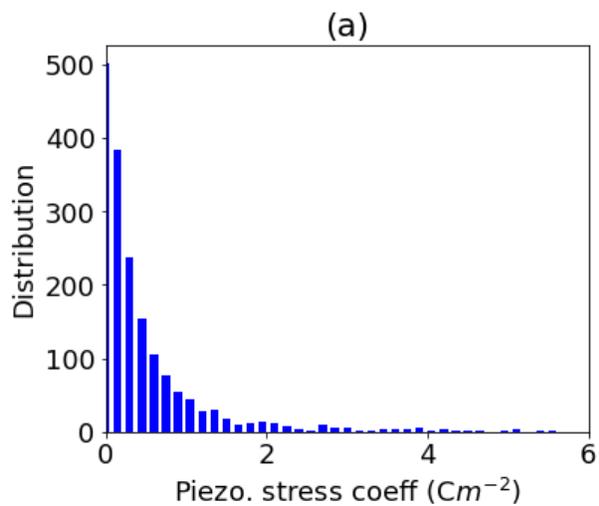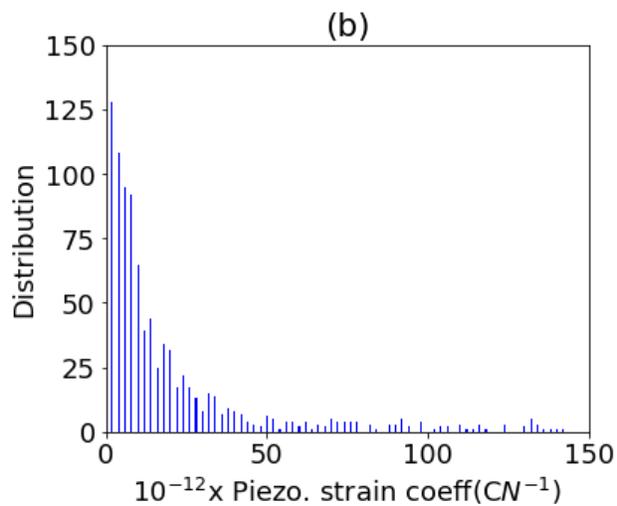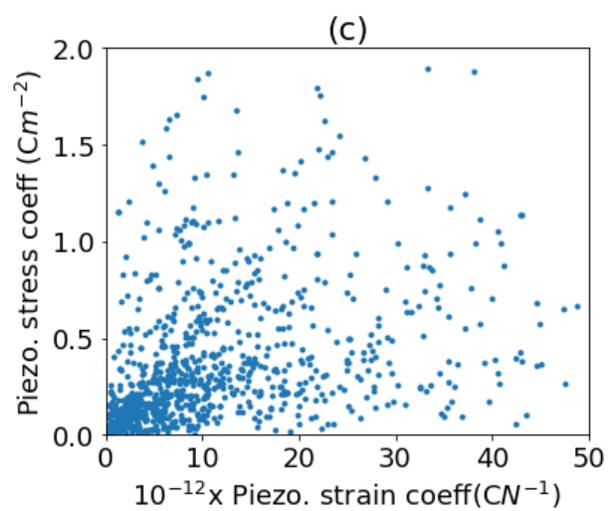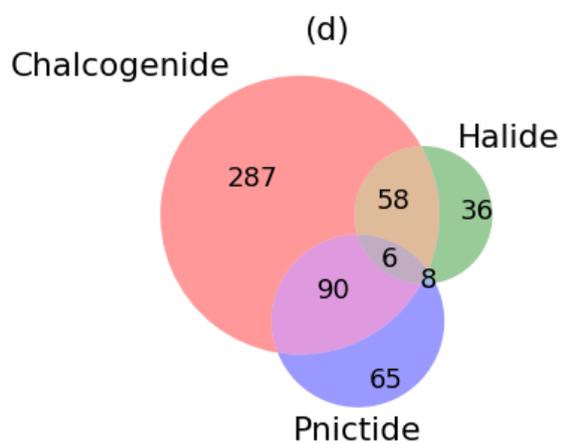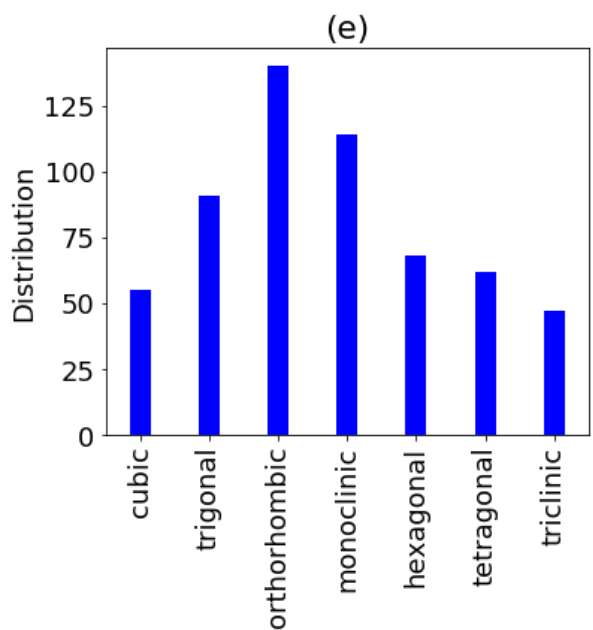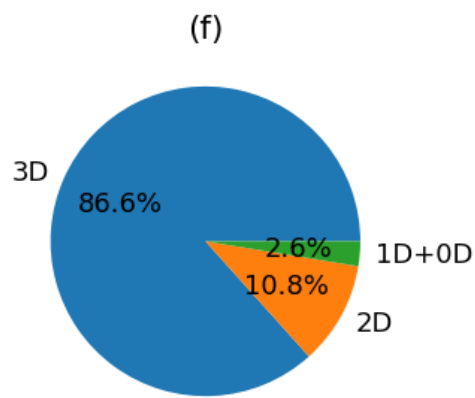

*Fig. 3 Piezoelectric data distribution. a) PZ stress coefficient distribution, b) PZ strain coefficient Distribution, c) PZ stress vs strain coefficients, d) chemistry of high-PZ materials, e) crystal system distribution of high PZ materials, f) dimensionality analysis of high PZ materials.*

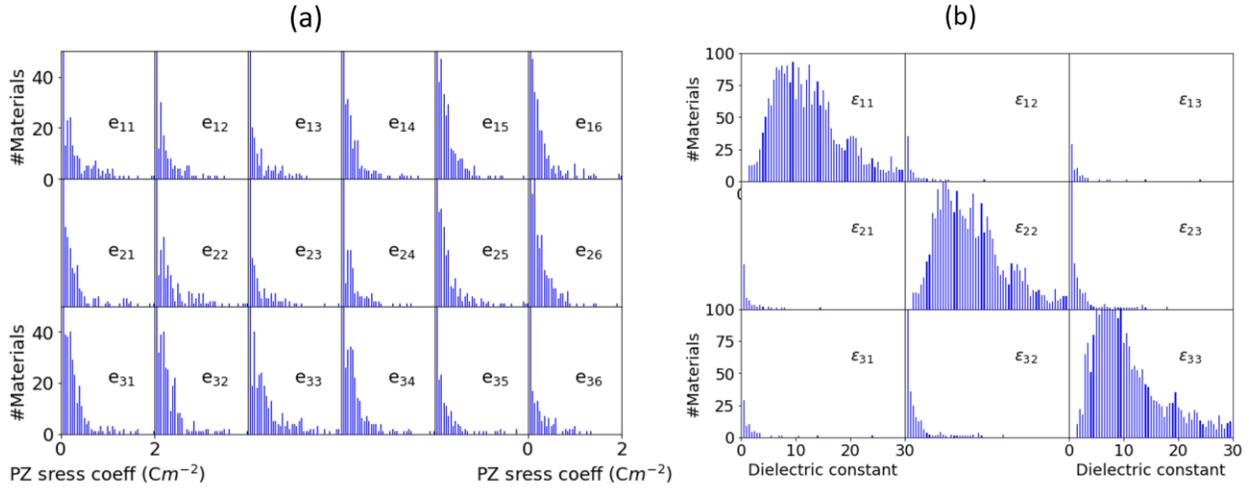

*Fig. 4 Dielectric and piezoelectric tensor distributions of the dataset considered in present work. a) the 3x6 piezoelectric tensor distribution, b) 3x3 dielectric tensor distribution.*

## 2.3 Dielectric properties

Dielectric materials are important components in many electronic devices such as capacitors, field-effect transistors computer memory (DRAM), sensors and communication circuits[59,60]. Both high and low-value DL materials have applications in different technologies. To evaluate the uncertainty in the data, we compare the dielectric constant of 16 materials with respect to experiments and the linear optics methods. The comparison is listed supporting information Table S3. We find the DFPT total DL has a mean-absolute deviation (MAD) of 2.46, which is lower than MAD of 2.78 for linear optics based Tran-Blaha modified Becke-Johnson (MBJ) potential with respect to experiments. Next, we analyze the dielectric constant data obtained using the DFPT method. In Fig. 5a, we plot the distribution of the average of the diagonal values of the dielectric tensor ($\varepsilon_{avg}$). We find a peak value of 8.9 and a long tail towards high values. Setting our threshold for high dielectric constant materials at 20, we identify 441 materials. Some examples of previously unreported low- $\varepsilon_{avg}$ materials are CClN( JVASP-14343), $GeF_4$( JVASP-22445) and $Mg(CN)_2$ (JVASP-29285), $BCl_3$ (JVASP-164), $PHF_4$ (JVASP-25550) while high-k materials are : $ZnAgF_3$ (JVASP-7792), $CaZrN_2$ (JVASP-79475), $Ta_2ZnO_6$ (JVASP-9231), $YWN_3$ (JVASP-38113), $KCaCl_3$ (JVASP-36962), $NbBi_3O_7$ (JVASP-13286), and RuC (JVASP-36402). In Fig. 5b, we compare the dielectric constants from this work to the low-frequency limit of the linear optics method we used to calculate the frequency-dependent dielectric function in our previous work[41]. The electronic part of the DFPT dielectric constant is in close agreement with that obtained from the linear optics method. Importantly, the ionic contribution to the static dielectric constant is frequently larger than the electronic contribution. Our dataset of electronic and ionic components can guide experimentalists to choose a material with either high and/or low ionic-contribution compounds. As expected, the electronic part of the dielectric constant and the bandgaps have a qualitative inverse relationship,[79] as shown in Fig. 5c. This relationship, combined with the underestimation of bandgaps by semilocal DFT functionals, can lead to an overestimation of electronic dielectric constants. Electronic bandgap properties are already discussed in detail in our previous work[41]. The dataset suggests that most of the high $\varepsilon_{avg}$ materials are 3D, with again a high number of chalcogenides (Fig. 5d). However, the low-dimensional dielectric materials could be of great technological importance, because the capacitance of a layer is generally inversely proportional to its thickness, potentially allowing for ultrathin vdW-bonded devices. Examples of 3D/2D/1D/0D bulk dimensional materials with a high and low-dielectric constant are shown in Table. 4 and Table. 5. We find similar crystal system trends for high DL and PZ materials. In Fig. 5e and 5f we show the crystal system and dimensionality trends of the screened materials. We observe that trigonal and tetragonal crystal systems are highly favored for the high DL materials. Similar to the PZ data, most of the high PZ value materials in our database are 3D. The electronic part of the dielectric constant can be directly used to estimate other physical properties such as the refractive index and the birefringence, which will be analyzed in detail in a follow-up work in the future.

*Table. 4 Some examples of materials with low crystallographic average dielectric constant ($\varepsilon_{avg}$). Complete data is available in the supplementary information.*

| Materials | Dimensionality | JID | Spg | $\varepsilon_{avg}$ |
|---|---|---|---|---|
| **CClN** | 3D | JVASP-14343 | 59 | 3.1 |
| **CSO** | 3D | JVASP-5482 | 160 | 2.8 |
| **SiO$_2$** | 3D | JVASP-54225 | 115 | 3.0 |
| **LiBF$_4$** | 3D | JVASP-21785 | 152 | 3.8 |
| **AlPO$_4$** | 3D | JVASP-4564 | 82 | 3.8 |
| **Mg(CN)$_2$** | 2D | JVASP-29285 | 102 | 2.92 |
| **Zn(CN)2** | 2D | JVASP-29282 | 102 | 3.03 |
| **SiS** | 2D | JVASP-28397 | 53 | 3.07 |
| **AgB(CN)$_4$** | 2D | JVASP-10675 | 215 | 3.2 |
| **ZnC$_2$S$_2$(OF)$_6$** | 2D | JVASP-29289 | 148 | 4.2 |
| **GeF$_4$** | 0D | JVASP-22445 | 217 | 3.2 |
| **BCl$_3$** | 0D | JVASP-164 | 176 | 2.60 |
| **PHF$_4$** | 0D | JVASP-25550 | 14 | 2.90 |
| **BH$_3$** | 0D | JVASP-33032 | 14 | 2.5 |
| **SiH$_4$** | 0D | JVASP-5281 | 14 | 2.8 |

*Table. 5 Some examples of materials with high crystallographic average dielectric constant ($\varepsilon_{avg}$). Complete data is available in the supplementary information.*

| Materials | Dimensionality | JID | Spg | $\varepsilon_{avg}$ |
|---|---|---|---|---|
| **ZnAgF$_3$** | 3D | JVASP-7792 | 221 | 92.3 |
| **CaZrN$_2$** | 3D | JVASP-79475 | 166 | 87.4 |
| **Ta$_2$ZnO$_6$** | 3D | JVASP-9231 | 44 | 74.9 |
| **KCaCl$_3$** | 3D | JVASP-36962 | 1 | 68.4 |
| **NbBi$_3$O$_7$** | 3D | JVASP-13286 | 1 | 62.8 |
| **PbO** | 2D | JVASP-29376 | 57 | 85.4 |
| **Bi$_2$TeSe$_2$** | 2D | JVASP-13955 | 166 | 80.7 |
| **WSeS** | 2D | JVASP-28903 | 164 | 76.5 |
| **CuSe$_2$Cl** | 2D | JVASP-5827 | 14 | 61.8 |
| **TiNCl** | 2D | JVASP-3894 | 59 | 60.0 |
| **Sb$_2$S$_3$** | 1D | JVASP-4285 | 62 | 70.0 |
| **SbSI** | 1D | JVASP-5197 | 59 | 58.8 |
| **Mn(SbS$_2$)$_2$** | 1D | JVASP-32156 | 12 | 57.0 |
| **SbSBr** | 1D | JVASP-5191 | 62 | 56.3 |
| **SbSeI** | 1D | JVASP-5194 | 62 | 49.0 |
| **Ta(TeBr$_3$)$_2$** | 0D | JVASP-5662 | 2 | 15.1 |
| **Pd(SeBr$_3$)$_2$** | 0D | JVASP-4080 | 2 | 12.1 |

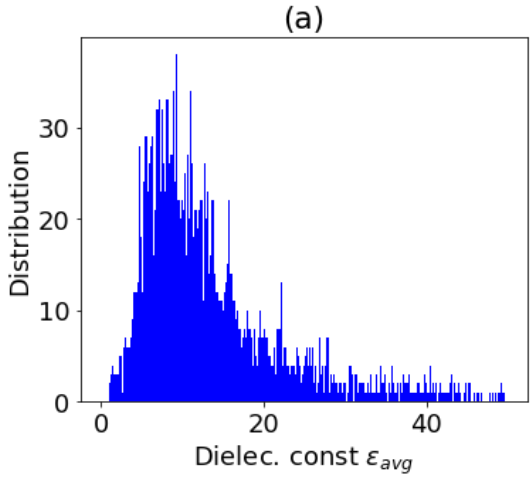
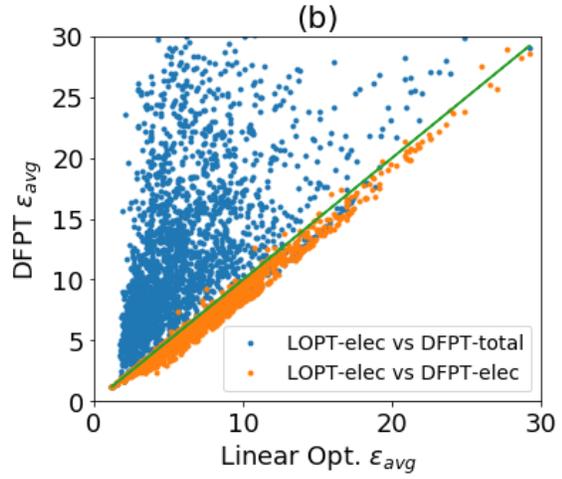
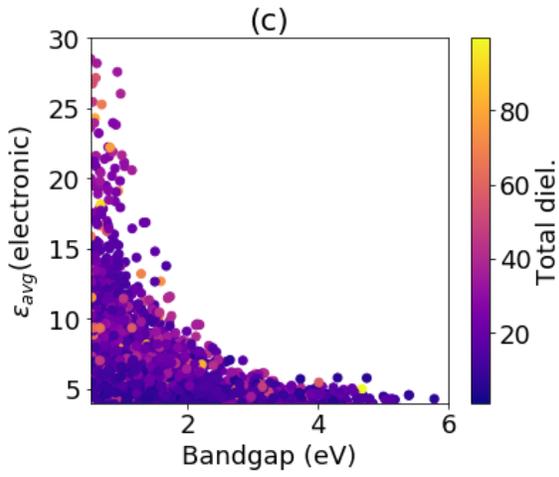
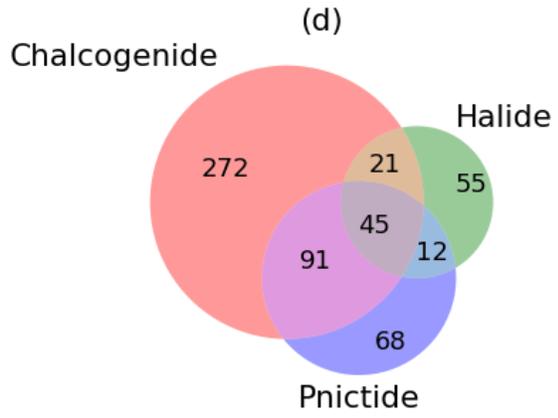
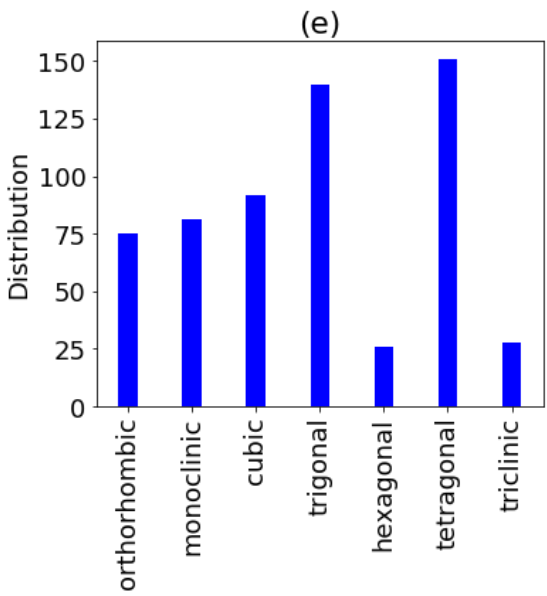
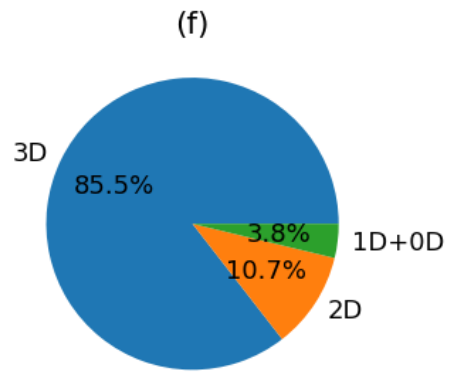

*Fig. 5 Analysis of the dielectric constant data a) dielectric constant distribution, b) dielectric constant comparison from linear optics and DFPT methods, c) dielectric constant vs bandgaps of materials, d) dimensionality analysis of high dielectric constant materials, e) crystal system distribution of high dielectric constant materials, f) dimensionality (3D/2D/1D/0D) analysis of high DL materials.*

## 2.4 Dynamically unstable materials

Our analysis finds many materials with imaginary phonon frequencies at Γ-point, which we call dynamically unstable materials, and which we exclude from our main analysis. In order to be observed experimentally, these materials must be stabilized by finite temperature contributions to the free energy, which are neglected in this analysis. Unstable phonon modes typically indicate that materials will go through phase transitions, and are therefore of interest for applications as ferroelectrics, antiferroelectrics, ferroelastics, etc. We find 1061 materials with at least one imaginary phonon frequency at Γ-point. We observed that most unstable materials have high crystal symmetry, and would likely go through symmetry-lowering phase transitions. Many of these systems belong to oxide families such as $ABO_3$ (examples: $BaTiO_3$ (JVASP-110), $KIO_3$ (JVASP-22568), $CoBiO_3$ (JVASP-29444), $ZrPbO_3$ (JVASP-7966), $MgZrO_3$ (JVASP-36637)), $AO_2$ (examples: $NbO_2$(JVASP-12003), $BiO_2$(JVASP-12066), $FeO_2$ (JVASP-18430) ). In addition to the oxide family, we also find several other classes, including halides, hydroxides, and chalcogenides. Related high-throughput searches of unstable materials have recently been carried out by Garrity[80]. A full analysis of dynamically unstable materials would require finding all of the low energy phases of an unstable material, as well as how they are related. This goes beyond the scope of this work, but may be the topic of future work. A list of dynamically unstable materials is given in the supplementary information (see data-availability section).

## 2.5 Born-effective charge analysis

In Figs. 6 and 7 we show some of the general trends in the PZ, IR and dielectric data. The relaxed ion contributions to all the properties considered in this work depend on the Born effective charges, making understanding trends in the BEC important to understanding the entire dataset. In the Fig. 6a the overall distribution of the maximum BEC in a material and in Fig. 6b the absolute difference in maximum BEC and the maximum formal charges[81] in a material. While most materials have BEC near their formal charges, there are significant outliers with anomalous BEC, which can represent some of the best candidate materials for piezoelectric and dielectrics and is commonly associated with ferroelectrics. For example, Figs. 6(c-d) portray the maximum BEC charge of a material and the minimum frequency versus the piezoelectric and ionic dielectric constant, and it is evident that almost all the high-response materials have high BEC and low-frequency modes at $\Gamma$-point. We note that while anomalous BECs are often a signature of ferroelectric materials, we remove materials with unstable phonon modes at $\Gamma$ in this work.

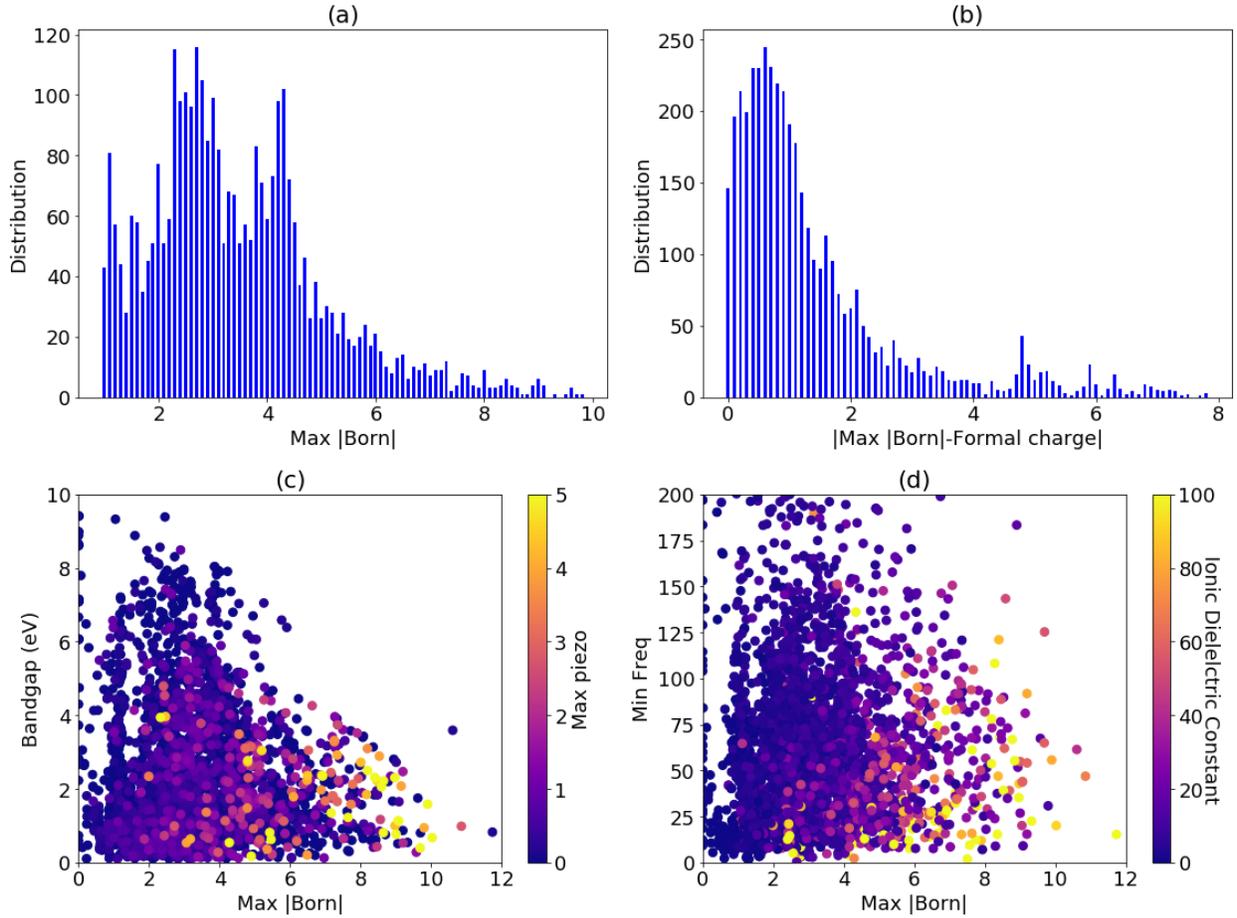

*Fig. 6 Born effective charge distribution and its relation to PZ, DL and bandgaps. a) Histogram of Born-effective charge data, b) histogram of maximum BEC and formal charge c) BEC wrt bandgap with color-coded max PZ, d) BEC with respect to the min-IR-frequency with color-coded ionic part of dielectric constant.*

## 2.6 Chemical Trends

Next, we depict the trends across the periodic table of the properties investigated in this work in Fig. 7. In order to understand the contribution of various elements to a given property, we weigh an element in a material one or zero depending on whether the material has a maximum-IR for far-IR peak (Fig. 7a), has PZ value greater than 0.5 $C/m^2$ (Fig. 7b), has dielectric constant more than 10 (Fig. 7c) and has BEC more than 5 or not (Fig. 7d). After such weighing for all the materials in our dataset, we calculate the probability that an element is part of a high-value material using the threshold defined above. For example, suppose there are x number of Se-containing materials and y of them have a property over the threshold, then the percentage probability ($p$) for Se is calculated as: $p = \frac{y}{x} \times 100\%$. We find that high-IR peaks generally have the light elements H, B, C, N, or O, as discussed earlier. The highest PZ, DL and BEC most commonly have Ti, Zr, Hf, Nb, Ta, or Bi. The first five elements are commonly found in 4+ and 5+ oxidation states in insulators, which tends to result in high BECs, and these elements are also present in many ferroelectric perovskite oxides. Most of the transition metals after $d^5$ have low IR/PZ/DL and BEC values that can be related to the partial-filling of d-orbitals. Similar behavior for other materials properties such as elastic properties has been observed[23]. Somewhat surprisingly, our database shows the Bi has a higher average piezoelectric response than neighboring lone-pair element Pb, which is commonly associated with a high-piezoelectric response, especially in PZT, suggesting that Bi may be effective as a Pb-free piezoelectric substitute.

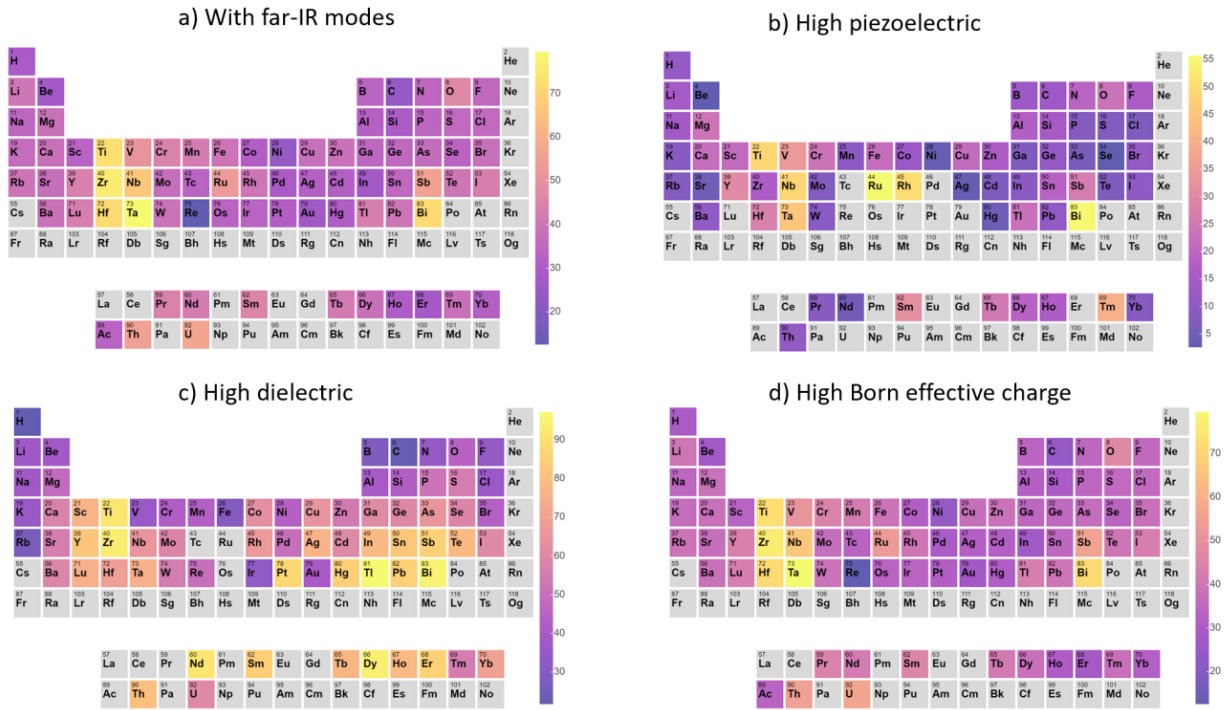

*Fig. 7 Trends across the Periodic table of a) high IR peak (having at least one peak>400 cm$^{-1}$), b) piezoelectric (>1 C/m$^2$), c) dielectric (>20) materials, and 2) max. Born-effective charge in a system (>5) The elements in a material are weighed 1 or 0 if the material has high or low-values. Then the percentage probability of finding the element in a high-value material is calculated.*

## 2.7 Machine learning

Recent advances in data infrastructure, statistics, machine learning, and computational methods has led to an explosion of computed data in the field of materials science[25,26,82-86]. Here, we apply the machine learning classification model using classical force-field inspired descriptors (CFID)[46], gradient-boosting decision trees (GBDT)[87,88] to our dataset consisting of 3954 dynamically stable materials. We partition the complete dataset of highest IR frequency, highest Born-effective charge in a system, highest PZ tensor value and average dielectric tensor in 90%-10% train-test divisions. To obtain optimized parameters (number of trees, number of leaves and learning rate) for each model, we carry out hyper-parameter optimization with five-fold cross-validation on the 90% training data for each model. First, we train a regression model for predicting the highest IR frequency mode of a material. The mean absolute error (MAE) and $r^2$ on the 10% held set for such a model are 67.8 cm$^{-1}$ and 0.96, respectively. A scatter plot of the performance on the held set is shown in Fig. 8a. While the MAE of the ML model of 67.8 cm$^{-1}$ is about 8 times higher than the MAE of DFPT with respect to experiments, it is small with respect to the whole range of investigated values. The high $r^2$ value and obvious trend in Fig. 8a shows that the maximum IR frequency can be at least roughly predicted using M. Furthermore, a five-fold cross-validation shows a MAE score of 77.84 ±34.67 cm$^{-1}$ supports such a conclusion. The corresponding learning curve (Fig. S1) shows that the model error decreases as we add more data, suggesting that this model will improve as our database increased.

GBDT inherently allows accessing the importance of all the features of the models. For this model, feature importance analysis shows that some of the most important descriptors are radial distribution function data, especially at 1.5 and 5.9 Å, electron affinity, and atomic radii of elements. The high importance of short bond-lengths can be easily understood as it indicates the presence of H-containing compounds, which tend to have high IR modes as discussed earlier.

Similarly, we train a regression model for maximum Born effective charge (BEC) in materials. Predicting for the test dataset, we find the MAE and $r^2$ in BEC as 0.6 and 0.76 respectively as shown in Fig. 8c, showing that a significant portion of the BEC can be predicted using ML. A five-fold cross-validation shows MAE of 0.69±0.31, therefore supporting the above conclusion. Given the non-trivial relationship between the BEC and the formal charge, shown in Fig. 8b, it may be difficult to improve this model. Some of the most important features are the radial distribution function at 3.4 Å, 6.2 Å, angle distribution at 143º, dihedral angle at 95º, atomic radii and heat of fusion of elements. We notice that the standard deviation for cross-validation in both the above two regression models is about half of the actual MAE values (IR:34.67 vs 77.84 cm$^{-1}$ and BEC: 0.31 vs 0.69). We provide the learning curves for these regression models in supporting information (see Fig. S1). Again, it shows that the models would improves if we add more training data.

We try similar regression models for a maximum value of the PZ tensor (MAE: 0.47 C/m$^2$) and average DL tensor (MAE: 4.91), but these regression models perform poorly, with $r^2 < 0.2$ in both cases. Therefore, we train classification models for predicting the high PZ and high dielectric models using a threshold of 1 C/m$^2$ for PZ and 10 for the dielectric constant, instead of trying to predict exact values, as for IR and BEC. In classification models, the accuracies are evaluated in terms of the area under curve (AUC) of the receiver operating characteristics (ROC) curves (Fig. 8b, d). The ROC curve illustrates the model's ability to differentiate between high and low-performance materials. We find ROC AUCs of 0.86 and 0.93 for the high PZ and high DL materials. For each property, the five-fold cross-validation result is shown with the gray region around the ROC curve. We find a very small deviation due to the swapping of datasets during cross-validation, indicating the model is fairly agnostic to the choice of test data. For the above models, in addition to the structural descriptors, chemical descriptors such as heat of fusion and maximum oxidation states of constituent elements are important. We note that the structural features are more important for the above DFPT based properties than quantities such as formation, exfoliation energies, bandgaps where chemical descriptors were more important as shown in our previous work[46]. This highlights the higher difficulty in predicting the DFPT-based properties studied in this work, as these properties depend strongly on the structure of the material, rather than mostly on just the type of elements and bonding.

All these models, along with the data, are publicly available at (https://www.ctcms.nist.gov/jarvisml/) to predict the performance of new compounds. These ML models can be used to identify materials worth performing the next set of DFPT calculations. For instance, we apply the classification models for high DL values on 1193972 materials taken from AFLOW[4], Materials-project[3], Open Quantum Materials Database (OQMD)[5], Crystallography Open Database (COD)[89] and JARVIS-DFT databases combined. We convert the structures from these databases into CFID descriptors and then we can easily apply the trained models. As the predictions in ML are very fast, we quickly pre-screen 32188 high DL materials. Now, these new materials can be prioritized in the next set of DFT calculations in our database workflow. We have applied such workflows previously for various quantities such as exfoliation energies, solar-cell efficiency, and thermoelectric with appreciable success.

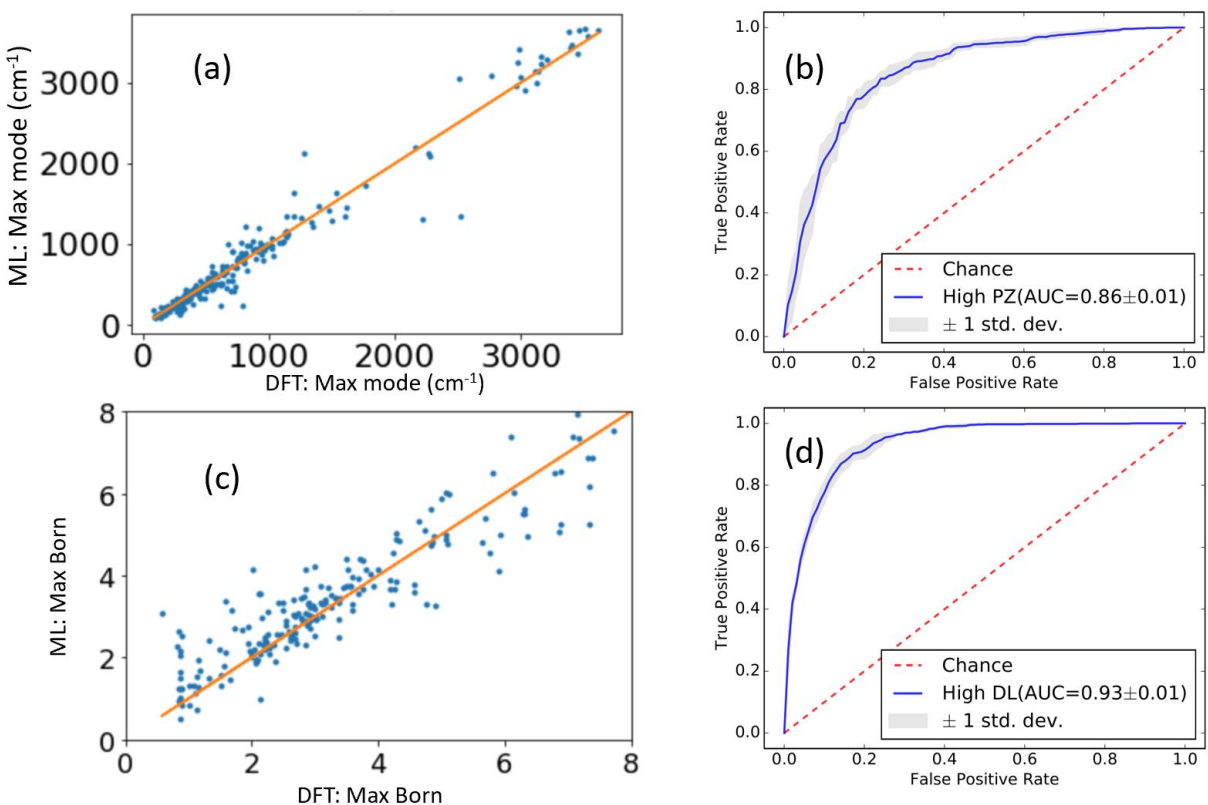

*Fig. 8 Machine learning regression model for predicting highest IR peak, b) classification model ROC curve for high PZ materials, c) regression model for maximum BEC in a material, d) ROC curve for high dielectric materials.*

## 3 Summary & conclusions

In summary, we perform a multi-step high-throughput computational screening study for infrared, piezoelectric and dielectric materials, using density functional perturbation theory on 5015 inorganic materials. These data constitute one of the largest datasets of infrared, dielectric and piezoelectric properties. Creating such a repository is a necessary step for data-driven material-design. We verify our workflow by comparing our computational results with several experimental measurements and alternative computational techniques like finite-differences, finding strong agreement. Using this database, we analyze the trends in these properties in terms of the dimensionality of materials (0D/1D/2D/3D), space-group, and chemical constituents, and we find various correlations that quantitively confirm and match some known chemical and physical trends and may help designing improved materials. We have identified several candidate compounds for high-performing infrared-detectors, piezoelectrics and dielectrics that have not been studied experimentally yet, to the best of our knowledge. To summarize, some of them are listed in Table 1-5. We observe that many hydroxides and halides have high and low-IR modes, respectively, as shown in Table 1 and 2, which are trends that can be used for the rational design of IR-detectors. We identify several candidates with high piezoelectric stress and strain coefficients, and notice that they are generally oxides and belong to orthorhombic crystal system. Somewhat surprisingly, we note that PZ-stress and PZ-strain coefficients are not strongly correlated. We verify the inverse relationship between bandgap and dielectric constants at a large scale and predict several new materials with unusually low and high constants (shown in Fig. 5c). We find that most of the halides have low while oxides have high dielectric constants. Using the phonon data at Γ-point, we identify many dynamically unstable materials, which may include new functional materials like ferroelectrics. Most of the high piezoelectric, dielectric and Born-effective charge materials contain elements such as Ti, Zr, Hf, Nb, Ta and Bi. Most of the high-performance materials in the above classes are three-dimensional (i.e. no vdW bonding) in nature, but there are also several candidates of 2D, 1D and 0D materials that can be useful for flexible and unique electronics applications. In addition to finding trends and correlations, we use our data to train machine learning surrogate models to expedite the computational screening processes. We achieve high accuracy models for predicting Born effective charges, maximum IR mode and classifying high piezoelectric and dielectric materials. We find bond-lengths, electron-affinity and atomic radii as some of the most important features in the machine learning models which are also intuitive and helpful for physical understanding. We believe that our results, workflow and tools can act as a guide for the experimental synthesis and characterization of various next generation materials.

## 4 Methods

The DFT calculations were carried out using the Vienna Ab-initio simulation package (VASP)[90,91]. The entire study was managed, monitored, and analyzed using the modular workflow, which we have made available on our github page (https://github.com/usnistgov/jarvis). Please note that commercial software is identified to specify procedures. Such identification does not imply recommendation by the National Institute of Standards and Technology. We use the projected augmented wave method[92,93] and OptB88vdW functional[94], which gives accurate lattice parameters for both and non-vdW (3D-bulk) solids[6,23]. In this work, a material is defined as low-dimensional if it contains vdW-bonding in one (2D-bulk), two (1D-bulk), or three (0D-bulk) crystallographic directions[23]. Both the internal atomic positions and the lattice constants are allowed to relax until the maximal residual Hellmann–Feynman forces on atoms are smaller than 0.001 eV Å$^{-1}$. The k-point mesh and plane-wave cut-off were converged for each materials using the automated procedure in the JARVIS-DFT[47]. We assume that achieving absolute convergence in energy is sufficient for obtaining reasonable DFPT results, and this assumption is supported by the agreement between frozen-phonon/finite-difference/finite-displacement method (FDM) and DFPT as well as linear-optics and DFPT results (discussed later). We also carry out K-point convergence for three materials: Si, AlN, MgF$_2$ as test cases to show that our converged K-points are sufficient to predict the DFPT related data as shown in supplementary information (Table. S4). We carry out the DFPT calculation on the standard conventional cell for each material. DFPT calculations, as implemented in the VASP code, were used to determine the Born effective charge tensors and the phonon eigenvectors were determined using phonopy code[75].

Given an insulating system with $N$ atoms per cell, with cell volume $\Omega_0$, atomic displacements $u_m$ (m=(1…3N)), homogenous strain $\eta_j$ (j=(1…6)), homogenous electric fields $\mathcal{E}_\alpha$ ($\alpha$=(x,y,z)), energy $E$, the force-constant matrix ($K_{mn}$), internal strain tensor ($\Lambda_{mj}$), the dielectric susceptibility ($\chi_{\alpha\beta}$), Born dynamical effective charge ($Z_{m\alpha}$), piezoelectric stress tensor ($e_{\alpha j}$) and piezoelectric strain tensor ($d_{\alpha j}$) are calculated as follows (SI units are used throughout)[65-69]:

$$K_{mn} = \Omega_0 \frac{\partial^2 E}{\partial u_m \partial u_n}\bigg|_{\varepsilon,\eta} \qquad (1)$$

$$Z_{m\alpha} = -\Omega_0 \frac{\partial^2 E}{\partial u_m \partial \mathcal{E}_\alpha}\bigg|_\eta \qquad (2)$$

$$\Lambda_{mj} = -\Omega_0 \frac{\partial^2 E}{\partial u_m \partial \eta_j}\bigg|_\varepsilon \qquad (3)$$

$$\chi_{\alpha\beta} == -\frac{\partial^2 E}{\partial \mathcal{E}_\alpha \partial \mathcal{E}_\beta}\Big|_{u,\eta} + \Omega_0^{-1} Z_{m\alpha}(K^{-1})_{mn} Z_{n\beta} \quad (4)$$

$$e_{\alpha j} = -\frac{\partial^2 E}{\partial \mathcal{E}_\alpha \partial \eta_j}\Big|_u + \Omega_0^{-1} Z_{m\alpha}(K^{-1})_{mn} \Lambda_{nj} \quad (5)$$

$$d_{\alpha j} = S_{jk}^{(\mathcal{E})} e_{\alpha k} \quad (6)$$

The dielectric constant can be derived from the dielectric susceptibility using:

$$\varepsilon_{\alpha\beta} = \varepsilon_0(\delta_{\alpha\beta} + \chi_{\alpha\beta}) \quad (7)$$

In Eq. 4 and 5, the first term represents the electronic contribution and the second term the ionic contribution for DL and PZ constants respectively.

The PZ is a 3x6 tensor, the DL 3x3 and the BEC $N$x3x3 tensor. The IR intensity of phonon modes is calculated using:

$$f(n) = \sum_\alpha \left| \sum_{s\beta} Z_{\alpha\beta}(s) e_\beta(s,n) \right|^2 \quad (8)$$

where $e_\beta(s,n)$ is the normalized vibrational eigenvector of the $n$th phonon mode of the $s^{th}$ atom in the unit cell, and α, β are the cartesian coordinates. $Z_{\alpha\beta}(s)$ is the Born effective charge tensor of $s^{th}$ atom (here we explicitly write both the cartesian indices of Z). These approaches are universal and have been already applied to various material classes. More details about the DFPT formalism can be found in elsewhere[65,66].

Our machine-learning models were trained using gradient boosting decision trees (GBDT)[87,88] and classical force-field inspired descriptors (CFID) descriptors[46] using a five-fold cross-validation grid search on the 90% training set. Using the best model found during the grid search, we test the model on the 10% held set and report the performance. The accuracy of the regression models and classification models were evaluated using mean absolute error and receiver operating characteristics (ROC) curves, respectively. The principal idea behind the GBDT algorithm is to build new base learners to be maximally correlated with the negative gradient of the loss function associated with the whole ensemble. The CFID approach gives a unique representation of a material using structural (such as radial, angle and dihedral distributions), chemical, and charge properties for a total of 1557 descriptors. We trained machine learning regression models to predict the highest IR frequency and maximum BEC of a material and classification models to predict whether a material has high PZ coefficient (>0.5 C/m$^2$) and dielectric constant (>20).

## 5 Data availability

The electronic structure data is available at the JARVIS-DFT website: https://www.ctcms.nist.gov/~knc6/JVASP.html and http://jarvis.nist.gov . The dataset is also available at the Figshare repository: https://doi.org/10.6084/m9.figshare.11916720 .

## 6 Contributions

KC and KG jointly developed the workflow. KC carried out the high-throughput DFT calculations. KC, KG and FT analyzed the DFT data with the help from VS. AB and AH helped in the experimental validation section. All contributed in writing the manuscript.

## 7 Competing interests

The authors declare no competing interests.

# Supplementary information:

# High-throughput Density Functional Perturbation Theory and Machine Learning Predictions of Infrared, Piezoelectric and Dielectric Responses

Kamal Choudhary[1], Kevin F. Garrity[1], Vinit Sharma[2,3], Adam Biacchi[4], Angela R. Hight Walker[4], Francesca Tavazza[1]

1 Materials Science and Engineering Division, National Institute of Standards and Technology, Gaithersburg, Maryland 20899, USA.

2 National Institute for Computational Sciences, Oak Ridge National Laboratory, Oak Ridge, TN 37831, USA.

3 Joint Institute for Computational Sciences, University of Tennessee, Knoxville, TN, 37996, USA.

4 Engineering Physics Division, National Institute of Standards and Technology, Gaithersburg, Maryland 20899, USA.

Table S1 Comparison of experimental and DFPT IR frequencies (cm$^{-1}$).

| Mats. | JID | DFPT | Experiment |
|---|---|---|---|
| **ZnO** | 1195 | 379, 410 | 389,413[1] |
| **AlN** | 39 | 600, 653 | 620,669[2] |
| **GaN** | 30 | 532 | 531[3,4] |
| **SnS** | 1109 | 93, 144, 178, 214 | 99, 145, 178, 220 [5] |
| **SnSe** | 299 | 72.6, 98.44, 125.01, 160.3 | 80,96,123, 150 [5] |
| **KMgF3** | 20882 | 160.0, 287.0, 470.8 | 168, 299, 458[6] |
| **LiNbO3** | 1240 | 145.0, 216.6 | 160, 220[7] |
| **GeS** | 2169 | 106.4, 196.2, 236.5, 253.0, 276.9 | 118, 201,238,258, 280[8] |
| **MAD** | | | 8.36 |

Table S2 Comparison of piezoelectric coefficient max($e_{ij}$) data for experiments and DFT. We take average values for the cases where the experimental data are in a range.

| Mats. | JID | Max($e_{ij}$) | DFT | Reference |
| --- | --- | --- | --- | --- |
| **BN** | 57695 | 1.55 | 1.15 | 9 |
| **AlN** | 39 | 1.39-1.55 | 1.39 | 10-12 |
| **ZnS** | 7648 | 0.38 | 0.13 | 9 |
| **ZnSe** | 8047 | 0.35 | 0.06 | 9 |
| **SiO$_2$** | 41 | 0.171 | 0.16 | 13,14 |
| **BaTiO$_3$** | 110 | 1.94-5.48 | 4.13 | 15,16 17,18 |
| **LiNbO$_3$** | 1240 | 1.33 | 1.59 | 19 |
| **GaSb** | 35711 | -0.07 | -0.102 | 9 |
| **PbTiO$_3$** | 3450 | 3.35-5.0 | 3.96 | 20 |
| **GaN** | 30 | 0.73 | 0.47 | 21 |
| **InN** | 1180 | 0.97 | 0.90 | 22 |
| **AlP** | 1327 | -0.1 | 0.004 | 9 |
| **AlAs** | 1372 | -0.16 | 0 | 9 |
| **AlSb** | 1408 | -0.13 | 0.06 | 9 |
| **ZnO** | 1195 | 1.00-1.55,0.89 | 1.10 | 23 |
| **BeO** | 20778 | 0.1 | 0.22 | 23 |
| **MAD** | | | 0.21 | |

Table. S3 Comparison of static dielectric constant for DFPT, MBJ and experiment. Experimental data were obtained from [24-28]. MBJ data were obtained from our optoelectronic property database[29].

| Materials | JID | DFPT | MBJ | Experiment |
|---|---|---|---|---|
| **MoS$_2$** | 54 | $\varepsilon_{11}$ =15.56 | $\varepsilon_{11}$=15.34 | $\varepsilon_{11}$=17.0 |
| **MoSe$_2$** | 57 | $\varepsilon_{11}$=16.90 | $\varepsilon_{11}$=16.53 | $\varepsilon_{11}$=18.0 |
| **MoTe$_2$** | 60 | $\varepsilon_{11}$=21.72 | $\varepsilon_{11}$=18.74 | $\varepsilon_{11}$=20.0 |
| **WS$_2$** | 72 | $\varepsilon_{11}$=13.91 | $\varepsilon_{11}$=13.95 | $\varepsilon_{11}$=11.5 |
| **WSe$_2$** | 75 | $\varepsilon_{11}$=15.21 | $\varepsilon_{11}$=14.32 | $\varepsilon_{11}$=11.7 |
| **SiC** | 182 | 7.10 | 6.01 | 6.552 |
| **AlP** | 1327 | 10.33 | 6.94 | 7.54 |
| **BN** | 17 | $\varepsilon_{11}$=4.75 | $\varepsilon_{11}$=3.72 | $\varepsilon_{11}$=5.06 |
| **BP** | 1312 | 9.03 | 7.94 | 11.0 |
| **GaP** | 1393 | 13.22 | 8.33 | 11.11 |
| **AlSb** | 1408 | 12.27 | 9.87 | 12.04 |
| **ZnS** | 1702 | 9.39 | 4.8 | 8.0 |
| **CdTe** | 23 | 19.59 | 6.54 | 10.6 |
| **HgTe** | 8041 | $\varepsilon_{11}$=29.44 | $\varepsilon_{11}$=11.22 | $\varepsilon_{11}$=20 |
| **ZnSiP$_2$** | 2376 | $\varepsilon_{11}$=12.44 | $\varepsilon_{11}$=8.56 | $\varepsilon_{11}$=11.15 |
| **ZnGeP$_2$** | 2355 | $\varepsilon_{11}$=14.75 | $\varepsilon_{11}$=9.02 | $\varepsilon_{11}$=15 |
| **MAE** | - | 2.46 | 2.78 | - |

*Table S4 K-point dependence of average dielectric constant, maximum piezoelectric coefficient, maximum and minimum optical phonon modes for Si (JVASP-1002), AlN (JVASP-39), MgF$_2$ (JVASP-20134). Equivalent K-points are represented in length (Å unit), mesh and per atom units.*

| Materials | KP-length(Å) | KP-mesh | KP-per atom | $\varepsilon_{avg}$ | Max($e_{ij}$) | Max mode (cm$^{-1}$) | Min mode (cm$^{-1}$) |
|---|---|---|---|---|---|---|---|
| Si | 35 | 6x6x6 | 1728 | 12.9 | 0.0 | 496.5 | 153.3 |
| Si | 50 | 9X9X9 | 5832 | 12.8 | 0.0 | 496.4 | 153.0 |
| Si | 65 | 12x12x12 | 13824 | 12.8 | 0.0 | 496.3 | 153.1 |
| AlN | 25 | 9x9x5 | 1620 | 8.6 | 1.39 | 713.9 | 240.8 |
| AlN | 40 | 15x15x8 | 7200 | 8.6 | 1.39 | 714.0 | 240.8 |
| AlN | 55 | 20x20x11 | 17600 | 8.6 | 1.40 | 713.9 | 240.8 |
| MgF$_2$ | 20 | 4x4x7 | 672 | 5.26 | 0.0 | 507.0 | 86.1 |
| MgF$_2$ | 35 | 8x8x11 | 4224 | 5.26 | 0.0 | 507.0 | 86.2 |
| MgF$_2$ | 50 | 11x11x16 | 11616 | 5.26 | 0.0 | 506.9 | 86.2 |

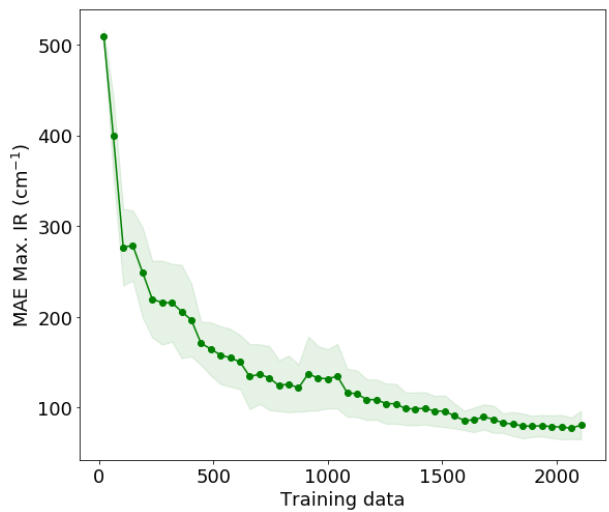
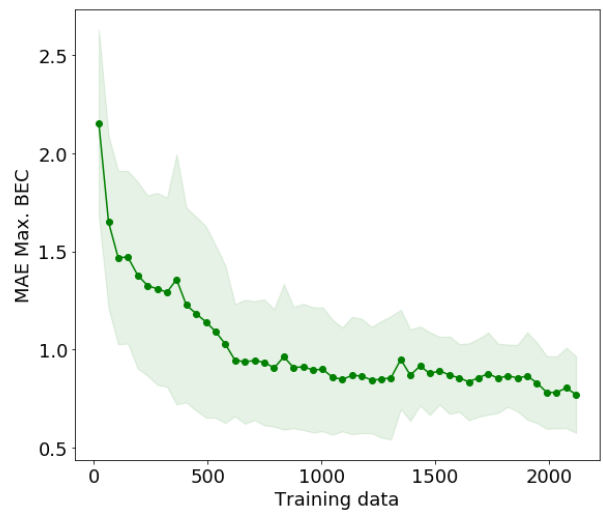

Fig S1 Regression learning curves for maximum IR mode and Born effective charge. The filled regions show standard deviation as we carry out five-fold cross-validation for each training data points in the figures.